# Proposal of Plane-Parallel Resonator Configuration for High-NA EUV Lithography

Tsumoru Shintake

Okinawa Institute of Science & Technology Graduate University OIST,

1919-1 Tancha, Onna, Okinawa, Japan 904-0495

shintake@oist.jp

Plane-parallel resonator configuration is proposed for high-NA EUV lithography, where the lithography mask and the wafer are parallelly arranged through two focusing mirrors. EUV light is injected through an off-axis rotating mirror at the back focal plane and provides off-axis illumination (precession beam) to the mask and bounces back twice (at the mask and the wafer), finally goes out from the resonator through the rotating mirror. This is a single path cavity, there is no resonant effect. The orbital error or vibration of the rotating mirror do not affect on the imaging quality. The off-axis illumination is essential for high-NA optics, which recovers the high spatial frequency, and improves the edge contrast. The matched annular-aperture is located at the back-focal plane of the projector mirror, which acts as Fourier filter passing only the horizontally scattered waves reflected by the density modulations $F(k_x, k_y, 0)$: 0-th z-order Fourier component of the mask. During single precession of the beam, this system creates 2D image of the normally projected density map of the mask pattern onto the wafer, where the longitudinal variation (3D effect) disappears, and thus the mask-shadowing problem is moderated. The depth-of-focus (DOF) is long, and also the image contrast becomes very high. As the objective mirror, Schwarzschild objective or Wolter telescope will be a suitable candidate. Wolter telescope is axisymmetric and lighter than conventional solid concave mirror, therefore, a larger diameter can be fabricated in high precession. By optimizing Wolter telescope, it may be possible to cover the wafer field size 26 mm x 33 mm of single patterning. Example parameter design was performed, $NA(\sin\theta) = 0.7$ seems technically feasible. If we remove the rotating mirror, the system becomes a resonator cavity, which will provide sensitive tool to verify errors in optical components and day-by-day corrections of mechanical alignments and also detecting mask defect.

# 1. Introduction

The precession method was developed in X-ray crystallography by Prof. Martin J. Buerger at the beginning of the 1940's as a very effective method to collect diffracted intensities without distorting the geometry of the reciprocal planes[1]. By rotating the crystal along X-ray axis(precession motion) and using annular aperture (slit) corresponding to intersection of Ewald's sphere and 0-th order plane, Bragg diffractions were collected. Recorded film showed the reciprocal planes without any distortion, thus it was used to determine the space group and unit cell dimensions, easily. The X-ray film acts as mathematical integrator summing up all diffractions during the rotation. Based on this concept, HUBER Buerger Precession Goniometer has been developed and widely used to analyze crystal structure, including protein crystals.

The same concept was realized in the electron microscope[2], where the electron beam rotates (precession motion) and illuminates on a stationary staying crystal. The precession method was also used to remove dynamical diffraction along zone axis[3], and could capture kinematic Bragg diffractions, which showed clean reciprocal lattice.

Without precession beam, we cannot capture all Bragg diffractions in one condition. Off-axis illumination enables coupling of wave and lattice at higher index (finer structure). Precession motion rotates the Ewald's sphere around the axis and covers all possible Bragg conditions. The annular aperture acts as Fourier filter and selects diffractions on 0-th order index, which is the Fourier transform of the projected density map of the crystal, or in our case it is the lithography mask.

In the Zernike's phase contrast optical microscope[4], the illuminating light is formed by annular aperture and focused by the condenser lens, forms the cone beam. At the back focal plane of the object lens, real image of the annular aperture is created, where the phase ring is placed to shift 90 degree of the optical phase. All diffractions from the sample are spreading wave (object wave) and reaches to the image plane (or CCD camera at the downstream), where the object wave interferes with the illuminating light (reference wave), and create interference fringe. By overlapping of all interference fringes, the real image is created. Because the bio-cell is "phase-object", i.e. a low contrast transparent object while cause phase shift due to index difference. Because the reference wave is phase shifted, the phase-object can be seen clearly with higher contrast.

The concept of Zernike's phase microscopy was introduced in lithography[5]. Using an annular-shaped phase filter in the projection optics, it was found to enlarge depth of focus for lines and spaces, isolated lines, spaces and hole patterns.

There is no precession beam in Zernike's microscope, while the cone beam is playing same role as the precession beam. Because the light source in the optical microscope is not transversely coherent, and incoming wavelets run through the annular aperture are fully independent, i.e., the optical phase and timing are randomly distributed. Photoemission at the light source is associated with energy transition of electrons, where those events are random. Therefore, the diffraction at the sample and followed by interference fringe creation, and photo-ionization process in the CCD detector are independent process of those incoming wavelets. By capturing image for a certain length of time, many number of events will be integrated on the detector, just same as, the precession beam does.

We have to be careful on FEL source at EUV wavelength[6]. It is transversely full coherent, and not suitable for lithography. Truncating beam at any aperture will cause strong interference pattern on the projected image on the wafer, which overlaps with the mask pattern. Light source with low transverse coherency is suitable for lithography.

The off-axis illumination has been successfully introduced in photo lithography[7], which enhances edge contrast. The lithography mask is not crystal structure, while the diffraction principle is same as X-ray crystallography, as we discuss in the later sections.

In case of EUV lithography, the off-axis illumination is effective. Unfortunately, there is a fundamental constraint: the mask itself is a mirror, thus the reflection angle has to be large enough to avoid overlapping of the incoming and outgoing beams. This situation introduces asymmetric optics in EUV lithography.

In this paper, we try to eliminate the constraint by introducing the **plane-parallel resonator configuration**.

## 2. Difficulties and possible solutions in EUV lithography optics

The key component toward narrow line width; the EUV light source at 13.5 nm is now available. Higher numerical aperture (NA) is still required, while there are following problems remained [9].

(1) 0.33 NA system is on production, and next generation high-NA (NA=0.55) system has been designed, which requires very large mirrors and extreme aspheres at the increased accuracy requirement. It is challenge to optics technology and manufacturing, resulting in higher cost.

Origin of this problem is that all optical components are used "off-axis", thus aspheres are required. We solve this problem by introducing a plane-parallel resonator configuration, where all optical components will be on-axis, and objective mirror is axisymmetric thus easy to fabricate in high precession.

(2) At high-NA, the contrast of the mask becomes low.

This is due to lower scattering amplitude at higher frequency. We solve this problem by introducing off-axis annular illumination (precession beam illumination), which recovers the high frequency diffraction amplitude, resulting in the higher contrast and sharper edge.

(3) Depth of focus is extremely short at high-NA, thus resist layer is not uniformly activated. We solve this problem by the annular aperture at the back-focal plane, which works as spatial filter, to select $F(k_x, k_y, 0)$ :Fourier component of projected image. The depth of focus becomes long, which is

$$DOF \approx \frac{\lambda}{2NA^2} \frac{R_a}{2w_a}$$

where $R_a, w_a$ is the radius and the width of the aperture, respectively. Typically, DOF becomes five to ten times longer than conventional system. Narrower width of the annular aperture will provide longer DOF. It must be optimized in experimentally, considering optical alignment errors and also the source size.

(4) Mask shadow effect becomes a fundamental problem in the high-NA lithography.
The plane-parallel resonator configuration moderates this problem. $F(k_x, k_y, 0)$ is projected density of the mask, where 3D effect disappears. We may also thinner the absorbing layer of the mask because of the higher contrast performance.

### 3. Proposed Optics for High-NA EUV Lithography

Figure 1 shows schematic illustration of the proposed EUV lithography having plane parallel resonator cavity configuration.
(1) The EUV light is injected through the rotating mirror near the back focal plane, which forms off-axis illumination (tilted angle) after the condenser mirror.
(2) At the mask, the illuminating beam is reflected back, and forms a spot focus through the projector mirror at the back-focal plane. The condenser and the projector are the same mirror. The position of the focus is opposite side of the rotating mirror, and thus, not blocked by the rotating mirror. It travels down through the objective mirror and illuminates the wafer surface as parallel beam, which acts as the reference wave for the beam interference.
(3) The reflected illumination beam is focused and travels back to the rotating mirror and goes out from the cavity.
(4) The diffracted beam from the mask is widely spreading beam, which carry the information of the mask structure (3D). When it arrives to the annular aperture, the central part is blocked. Those diffraction components are associated with the de-focus images and cause contrast degradation. The diffractions near the circle of the same radius as the off-axis illumination contain the projected image information, thus utilized as image formation through the annular aperture. It acts as spatial Fourier filter.

(5) The illuminating beam (reference wave) and diffractions (object wave) interfere each other and create interference fringes inside resist layer of the mask, and cause ionization at bright zone, as shown in Fig. 2.

(6) After single turn of the mirror rotation, the projected image is recorded inside the resist layer. Because the depth of focus (DOF) is long, the recorded image is uniform in longitudinally (except short periodic modulation due to interference effect between incoming and reflected illuminating beam).

(7) The overlapping de-focus component is weak, and thus the image becomes high contrast.

(8) As the objective mirror, Schwarzschild Objective or Wolter telescopes (type-I or type-III) will be a suitable candidate. Wolter telescope is axisymmetric and lighter than conventional solid concave mirror, therefore, a larger diameter mirror can be fabricated in high precession. Wolter mirror cancels coma-aberration by combining hyperboloid and paraboloid mirrors, however further optimization will be required in this application of much larger divergence compared to X-ray telescope. By using larger diameter mirror and optimizing curvatures, it may be possible to cover the wafer field size 26 mm x 33 mm.

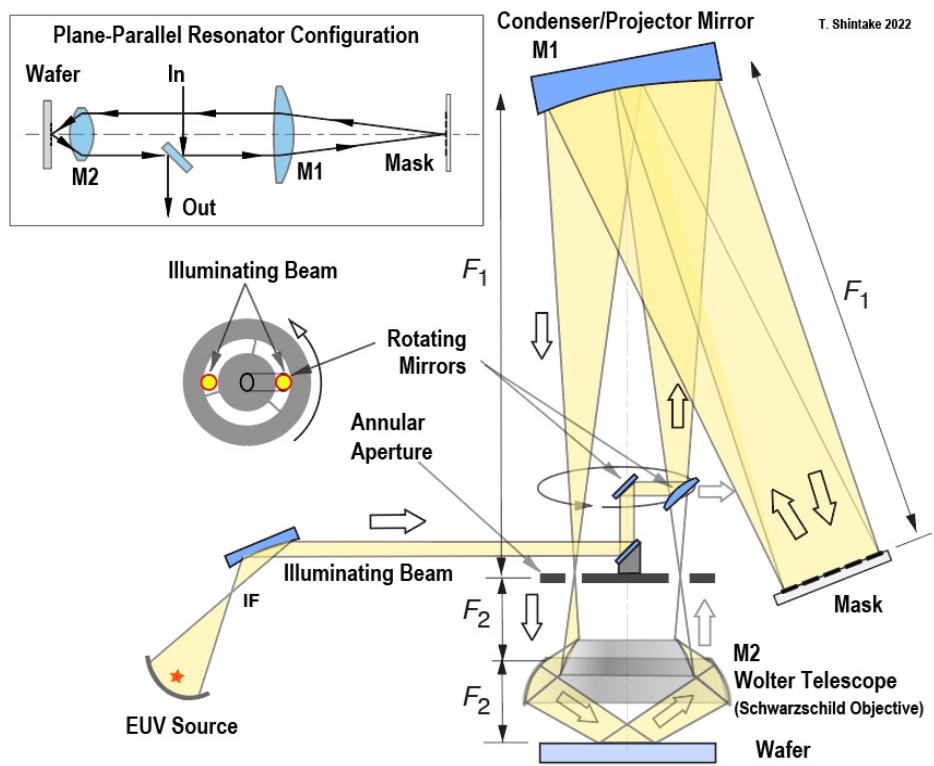

Fig. 1. Proposed high-NA EUV lithography optics. Only the illuminating beam is shown.

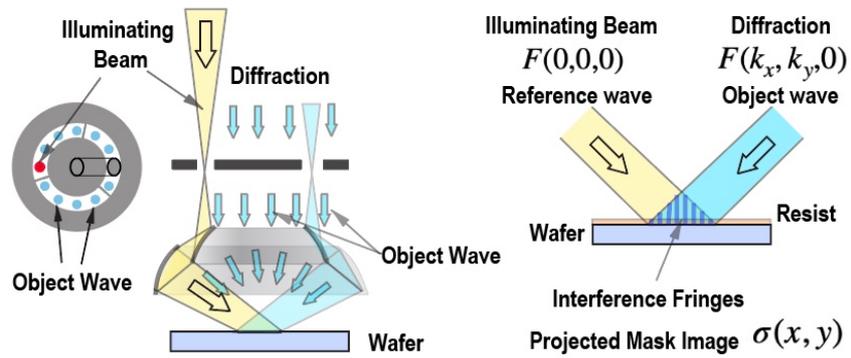

Fig. 2　Diffraction waves propagate down through the annular aperture, and reach to the wafer surface, where create interference fringes. During single mirror rotation, the projected image of the mask is reconstructed. Illuminating beam goes through the aperture, thus loss of the beam power is small. For more detail, refer to the later section theoretical background.

## 4. Case Study of Proposed EUV Lithography Optics

Currently running advanced EUV lithography has NA=0.33 maximum. The next generation NA=0.55 machine is under development, where the optics components are under test production[9]. Here we try to design, future generation NA=0.71 and NA=0.5 lithography optics based on proposed scheme. The following table shows results of estimation, they seem technically feasible. The depth of focus is much longer, roughly x5 times longer than conventional system.

Example Design Parameter

| Numerical aperture | NA | 0.71 | 0.5 |
|---|---|---|---|
| Wafer field size | $\Delta x, \Delta y$ | 26 mm x 33 mm | = |
| Magnification factor | M | x 4 | = |
| Mask size (diameter) | $M\Delta x, M\Delta y$ | 104 mm x 132 mm (168 mm) | = |
| Focal length of projector | $F_1$ | 1000 mm | = |
| Focal length of Wolter telescope | $F_2$ | 250 mm | = |
| Maximum beam angle at exit of Wolter telescope | $\theta_2$ <br> $NA = \sin\theta$ | 45 degree | 30 degree |
| Beam divergence at Wolter telescope | $\theta_{div} = \pm\frac{1}{2}\Delta y/F_2$ | ±3.8 degree | ±3.8 degree |
| Maximum diffraction angle at mask | $\theta_1$ <br> $\theta_1 = \theta_2/M$ | 11.3 degree | 7.5 degree |
| Tilt angle of precession beam on the mask | $\theta_T = 0.9 \times \theta_1$ | 10.1 degrees | 6.75 degree |
| Wolter telescope diameter at entrance | $D_W = 2F_1\theta_T$ | 350 mm | 235 mm |
| Annular aperture | $R_a = D_w/2 = F_1\theta_T$ <br> $w_a = 0.1 \times R_a$ | 175 mm <br> 18 mm | 118 mm <br> 12 mm |
| Projector mirror size | $D_p = 2F_1\theta_1 + D_{MSK}$ | 560 mm | 431 mm |
| **Depth of focus** | $DOF \approx \frac{\lambda}{2NA^2}\frac{R_a}{2w_a}$ | 67 nm | 135 nm |
| Mirror Rotation Speed | | 20 cycle/sec | = |
| EUV shots per cycle | 20 kH EUV Source | 1000 shots | = |
| Throughput*1 | at 100 stamps/wafer | 180 WPH | = |

Note (*1) Required photons dose has not yet been estimated. 2 cycles for one field image. 100 msec for stepper motion.

## 5. Proposed Optics for High-NA Optical Lithography at UV Wavelength

   Instead of rotating mirror, we can use annular aperture at upstream illumination system, this is similar to Zernike's phase microscopy, while working principle at downstream is unique in the proposed lithography.

(1) At this wavelength, we may use optical lenses, and all components are on-axis.
(2) The annular aperture in the upstream illumination system creates hollow beam, which provides off-axis illumination on the lithography mask, improves resolution.
(3) The illumination beam and diffracted beam travel down through projector lens, and filtered by the second annular aperture. The diameter of two apertures is identical, thus illumination beam goes through the aperture, and finally create mask image.
(4) The second annular aperture selects transversely scattered diffractions associated with the longitudinally projected mask image, and cut the de-focus components.
(5) The created image is high-contrast and DOF (depth of focus) becomes long.
(6) Diameter of the aperture should be smaller than the numerical aperture of the objective lens. Best results will be given at the outer edge is close to maximum aperture of the objective lens.
(7) Narrower width of the annular aperture will provide longer DOF. It must be optimized in experimentally, considering optical alignment errors and also the source size.

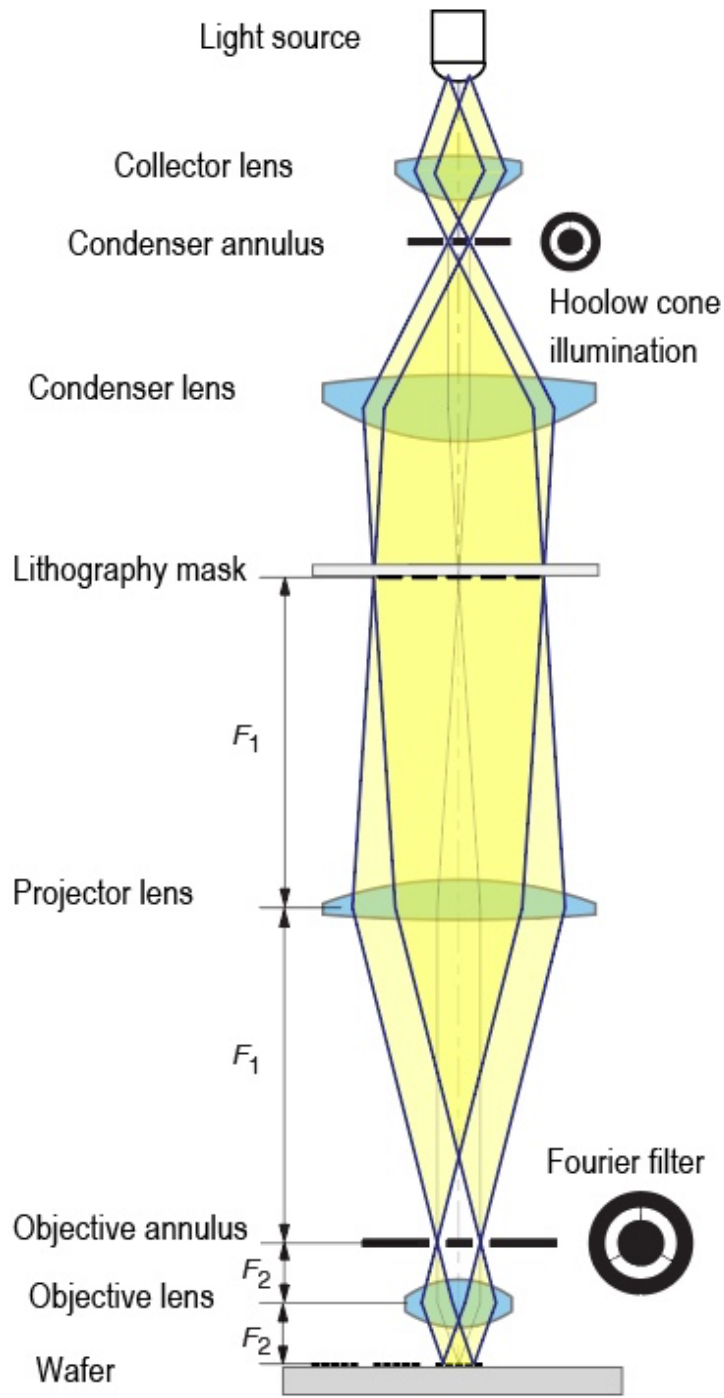

Fig. 3. Proposed optical lithography at UV wavelength. There is no rotating mirror, and all light waves propagate downward. If we fold the diagram in half at the lithography mask, it becomes the plane-parallel resonator configuration, shown in Fig. 1.

## 6. Theoretical Background

### 6.1 Conceptual 1:1 imaging system

We consider 1:1 image projection system as shown in Fig. 4, where two lenses of the same focal length-F are arranged in series with separation equal to twice focal length: 2F. If the illuminating beam is coherent, we have Fourier transform of the object at the back focal plane. We choose the diffraction beam at $(k_x, k_y)$ and also the central beam by two-hole aperture. After traveling down through the lens, those two beams will cross each other and create interference fringe around the image plane.

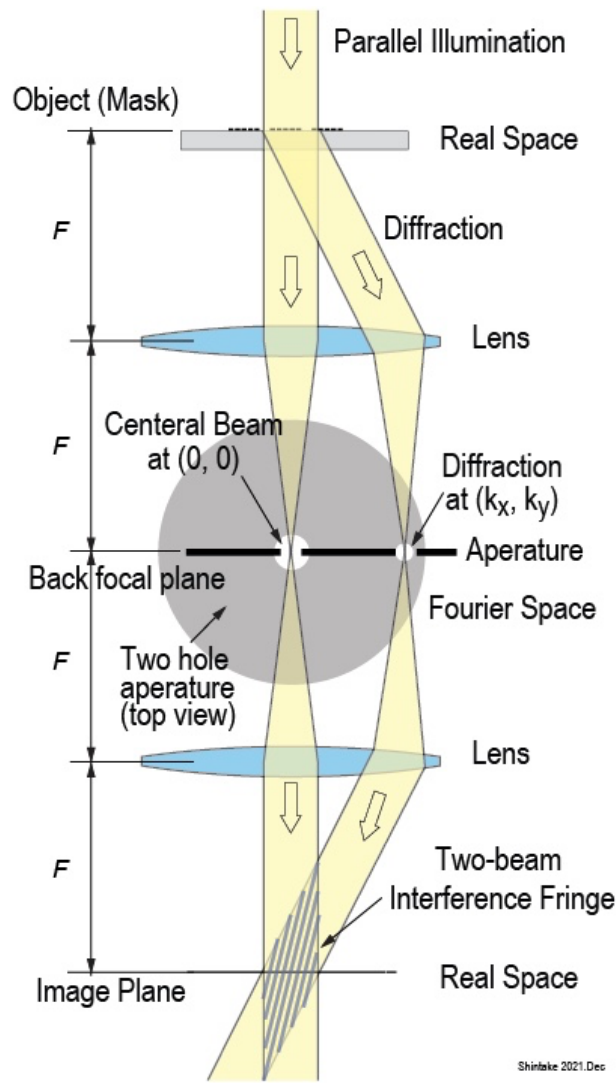

Fig. 4. Conceptual 1:1 imaging system. Diffraction beam at $(k_x, k_y)$ at the focal plan and the central beam are selected by two-hole aperture located at the back-focal plane.

In the next step, we propagate back the waves to the object plane. We may consider reversing the time clock. We should have the interference fringe around the object with the same pitch, i.e., which is a mirror image of the downstream interference fringe as shown in Fig. 5.

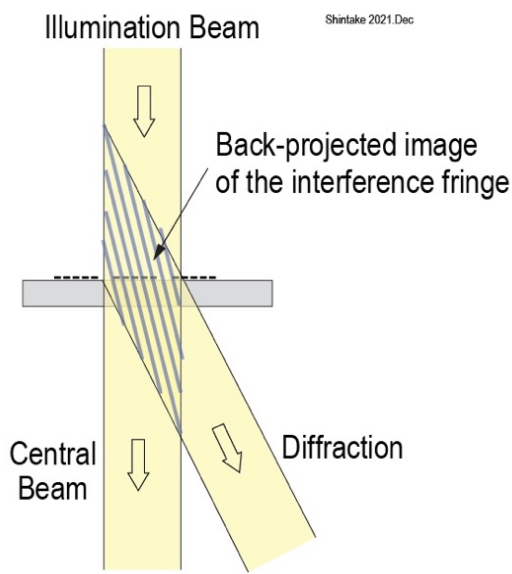

Fig. 5. Back projected image of the interference fringe.

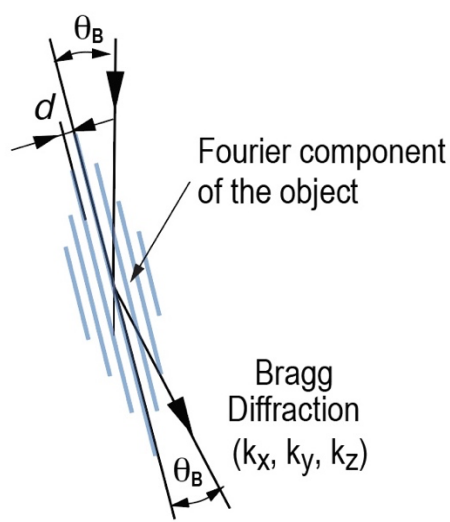

Fig. 6. Bragg diffraction with the periodic electron density modulation.

In the context of X-ray crystallography [9], the interference fringe is identical to the Fourier component of electron density of the object, which causes Bragg diffraction as shown Fig. 6. The reflection occurs when the following Bragg condition is satisfied.

$$2d \sin\theta_B = n\lambda \qquad (1)$$

where $\lambda$ is the wavelength of the light, $d$ is the spacing of the density modulation and $\theta_B$ is the reflection angle, *n* is integer. This phenomenon is also utilized as multi-coating mirror design in the EUV lithography system.

The reflectivity or the reflection beam intensity from the electron density modulation is proportional to the Fourier amplitude of the object at the scattering vector **S**.

$$\mathbf{S} = \mathbf{k}_2 - \mathbf{k}_1 \tag{2}$$

where $\mathbf{k}_1$ and $\mathbf{k}_2$ are the illuminating wave vector and the scattered (diffracted) wave vector, respectively. The Fourier component is given by

$$\mathbf{F(S)} = \int \rho(\mathbf{r}) \cdot \exp(i\mathbf{S} \cdot \mathbf{r}) \, dv \tag{3}$$

$\rho(\mathbf{r})$ is the electron density of the object, i.e., the lithography mask.

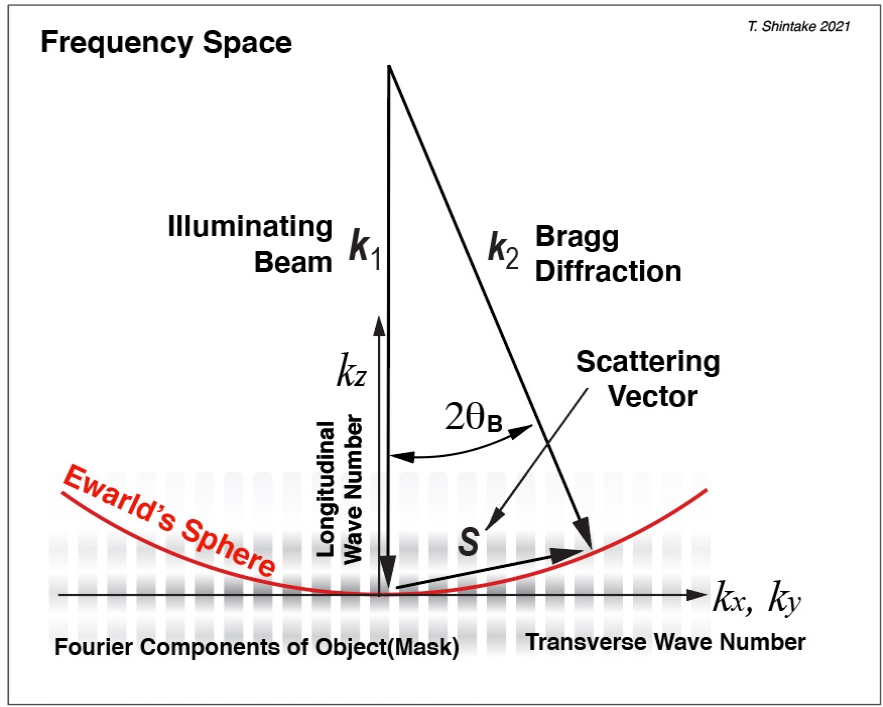

Fig. 7. Scattering vector **S**, Ewald's sphere and the Fourier component of the object.

The vector length of the wavenumber of scattered beam is same as the illuminating beam, thus those vectors should be on a sphere of radius $k_0$, which is called Ewald's sphere.

$$k_0 = 2\pi/\lambda = |\mathbf{k}_1| = |\mathbf{k}_2| = \sqrt{k_x^2 + k_y^2 + k_z^2} \tag{4}$$

The endpoint of the scattering vector picks up the amplitude of Fourier components $\mathbf{F(S)}$, which contains the phase, i.e., it is complex amplitude: $\mathbf{F} = |\mathrm{F}| \cdot \exp[i\alpha]$. Note that $\mathbf{F}$ is not a vector.

### 6.2 Improving Resolution and Contrast

When the lithography mask is illuminated by the coherent light from its normal direction as discussed in Fig. 7, there is the soften-edge problem due to high frequency loss in the projected image. As shown in Fig. 7, the scattering vector will follow on Ewald's sphere, and the longitudinal wavenumber $k_z$ becomes higher at higher frequency (higher scattering angle at higher resolution), where Fourier component is low, thus edge contrast becomes lower.

The transverse distribution of $\mathbf{F(S)}$ represents the reticle pattern on the mask, while the longitudinal distribution is related to the thickness of absorbing layer and its mirror image. $\mathbf{F(S)}$ can be written down in Cartesian coordinate as follows [5].

$$F(S_x, S_y, S_z) = \iiint e^{i2\pi(S_x x + S_y y + S_z z)} \rho(x, y, z) \, dx \, dy \, dz$$

$$= \iiint \rho(x, y, z) \, e^{i2\pi(S_x x + S_y y)} dx \, dy \cdot e^{i2\pi S_z z} dz \qquad (5)$$

Effect of the longitudinal structure variation is given by the last term. Thinner the reticle will provide broad spread of $\mathbf{F(S)}$ in longitudinal direction, and thus high frequency loss becomes less important. While in EUV lithography, to achieve enough contrast, the absorbing layer becomes longer than the wavelength at 13.5 nm, this effect is not negligible. Additionally, the mask absorbing pattern is created on the multilayer coating mirror, as a result, there exists double layer absorber images (real and mirror images), which cause shadowing effect and further complicated longitudinal Fourier components.

If the depth of focus is limited, the out of focus components will be blurred and overlap to the focused image, as a result, lower the contrast. To improve resolution and contrast, we need to obtain projected image of the mask, which is given by Fourier component on 0-th z-order plane [9].

$$F(S_x, S_y, 0) = \iint \left[ \int \rho(x, y, z) \, dz \right] \cdot e^{i2\pi(S_x x + S_y y)} dx \, dy \qquad (6)$$

Finally, we have projected image reconstructed at the wafer as follows.

$$\sigma(x, y) = \iint F(S_x, S_y, 0) \cdot e^{-i2\pi(k_x x + k_y y)} dk_x \, dk_y \qquad (7)$$

In this integration, $F(0,0,0)$ is the illuminating beam, which acts as the reference wave. $F(k_x, k_y, 0)$'s are the object waves interfere with $F(0,0,0)$, create interference fringes. Each fringe planes are normal to the wafer surface as shown in Fig. 2.

### 6.3 Off-axis annular illumination

Similar to the dark field optical microscope, and Zernike's phase contrast microscopy, off-axis illumination was employed at upstream of the condenser lens in the lithography, which enhances the resolution [2]. Basic concept of off-axis illumination is shown in Fig. 8. The illuminating beam is tilted by half of total scattering angle. The scattering vector **S** becomes close to 0-th order z-plane (horizontal), thus the higher frequency components will be recovered.

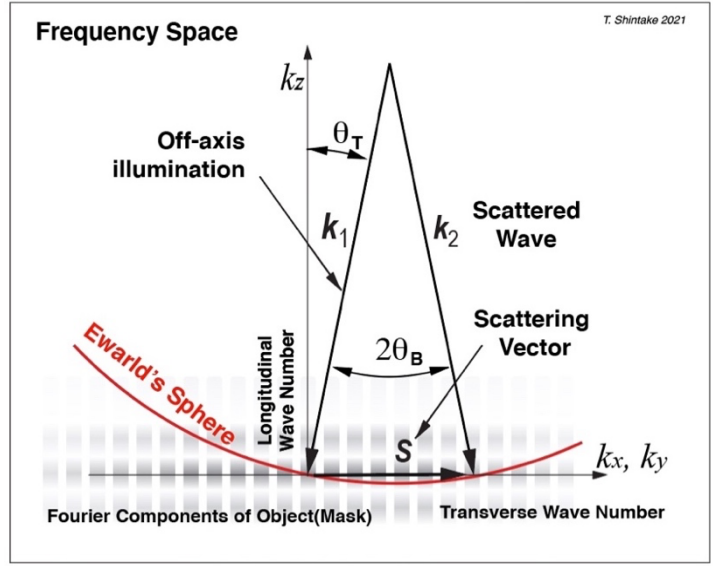

Fig. 8. Wave scattering with the off-axis illumination. The scattering vector **S** becomes closer to the 0-th z-order plane ($k_z \sim 0$).

As shown in Fig. 9, the intersection of Ewald's sphere to 0-th z-order plane becomes a circle, here we call "Ewald circle" in this paper. By rotating the illumination beam around the axis, Ewald circle rotates on the frequency space and thus covers all diffraction spots. Rotation steps has to be fine enough to avoid missing Bragg diffractions, associated with repeated reticle structure on the mask. The width of the circle is identical to the width of the annular aperture, which is roughly 10% of radius, thus we need 60 steps or more ($\geq 2\pi R_a / w_a$).

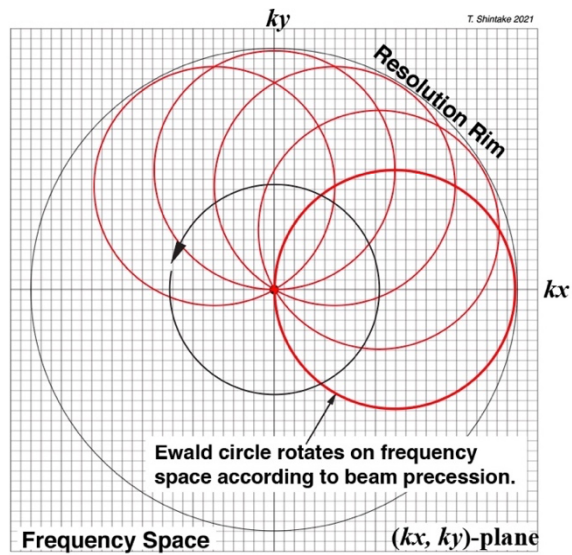

Fig. 9   On the frequency space, Ewald circle rotates according to the beam precession, and thus all diffraction spots are covered uniformly.

### 6.4. Introducing annular aperture

We introduce Fourier filter, i.e., the annular aperture in the projection optics. Fig. 10 shows perspective view of the wave scattering in real space. The scattering wave into Ewald circle is associated with the horizontal scatterings, which contain projected image information: $F(k_x, k_y, 0)$.

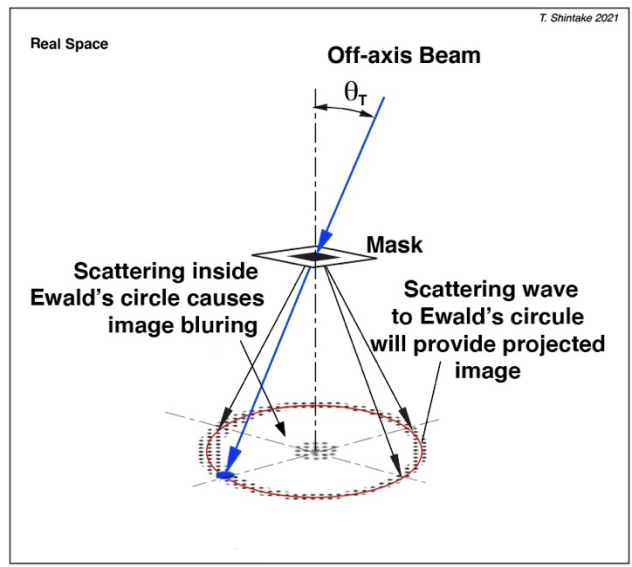

Fig. 10. Perspective view of the wave scattering. The scattering wave into Ewald circle is due to the horizontal scatterings, which contain projected image information: $F(k_x, k_y, 0)$.

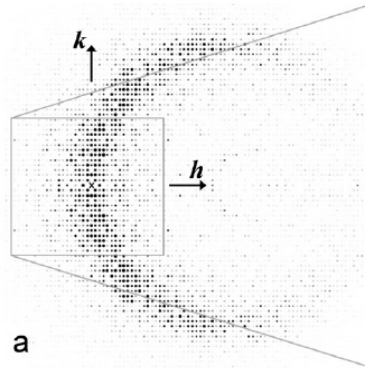

Fig. 11. Electron diffraction data with tilted beam illumination (copied from ref. 3), which clearly shows Ewald's circle.

In the beam precession study on electron diffraction crystallography, it was found the same effect experimentally as shown in Fig. 11 [3].

We introduce an annular aperture to select the diffraction of 0-th z-order, i.e., $F(k_x, k_y, 0)$, associated with projection image. This method was originally devised by Professor Martin J. Buerger to select Bragg diffractions of 0-th order, which was successfully used in the crystallography to understand crystal structure, started in 1940's [1].

After the annular aperture, the illumination beam and selected diffractions will interfere each other on the image plane. During single rotation of the mirror, illumination beam scans all azimuthal direction, and finally the projected density is recorded in the resist layer on the wafer, which is mathematically eq. (7).

# 7. Discussions and Conclusions

Working principle of Bragg diffraction and its applications are basically same concept as optical imaging on non-crystal structure, except the crystal structure has translational symmetry, which requests additional modification on Fourier amplitude, called "convolution integral"; well established mathematical procedure. Non crystalline structure produces smoothly distributed diffractions, and thus it is straightforward to process data and obtain images. The rotating mirror method is equivalent to rotation of crystal in X-ray crystallography. X-ray crystal request very careful angle scanning, since number of molecules in 3D crystal is huge, and Bragg diffraction condition is tight.

Lithography mask is similarly complicated huge system. Fortunately, it is still two-dimensional system, or in other words, we are interested in 2D structure, not 3D. Therefore, proposed method in this system would safely work to reduce complicity from 3D to 2D structure.

If we start 3D chip development, we have to struggle with huge amount of data in the future.

2007/4/5), ISBN 978-0387333342

[10] On EUV lithography.
  "EUV lithography systems", ASML Co. Website,
https://www.asml.com/en/products/euv-lithography-systems
"Extreme UV Photo Lithography", Newport Co. Website,
https://www.newport.com/n/extreme-uv-photolithography
"Extreme ultraviolet lithography" Wikipedia,
https://en.wikipedia.org/wiki/Extreme_ultraviolet_lithography